\documentclass[aps,prd,reprint]{revtex4-2}
\usepackage{xcolor}
\usepackage{hyperref}
\usepackage{amssymb}
\usepackage{amsmath}
\usepackage{graphicx}
\usepackage{xcolor}

\begin{document}
\title{Improved Constraints on the Minimum Length with a Macroscopic Low Loss Phonon Cavity}
\author{William M. Campbell}
\email{william.campbell@uwa.edu.au}
\author{Michael E. Tobar}
\affiliation{Quantum Technologies and Dark Matter Labs, Department of Physics, University of Western Australia, 35 Stirling Highway, Crawley, WA 6009, Australia.}
\author{Serge Galliou}
\affiliation{SUPMICROTECH-ENSMM, CNRS, institut FEMTO-ST, 26 Rue de l’Épitaphe 25000 Besançon, France}
\author{Maxim Goryachev}
\email{maxim.goryachev@uwa.edu.au}
\affiliation{Quantum Technologies and Dark Matter Labs, Department of Physics, University of Western Australia, 35 Stirling Highway, Crawley, WA 6009, Australia.}
\date{\today}
\begin{abstract}
Many theories that attempt to formulate a quantum description of gravity suggest the existence of a fundamental minimum length scale. A popular method for incorporating this minimum length is through a modification of the Heisenberg uncertainty principle known as the generalised uncertainty principle (GUP). Experimental tests of the GUP applied to composite systems can be performed by searching for the induced frequency perturbations of the modes of mechanical resonators, thus constraining the degree of minimum length in certain scenarios. In this work previous constraints made with mechanical resonators are improved upon by three orders of magnitude, via the utilisation of a cryogenic quartz bulk acoustic wave resonator. As well as purely mechanical resonant modes; hybrid electromechanical anti-resonant modes are investigated, and shown to be sensitive to the same GUP induced effects.
\end{abstract}
\maketitle
\section{Introduction}
The existence of a fundamental minimum length scale, on the order of the Planck length $l_p=1.62\times 10^{-35}$ m, is a prevalent idea in many theoretical models that contain a quantum description of gravity \cite{Luis1995, Hossenfelder2013}. Such a concept can be formulated in the context of non-relativistic quantum mechanics through modifications of the Heisenberg uncertainty relations. So called \textit{generalised uncertainty principles} \cite{Maggiore1993, Amati1989, Jizba2010, Scardigli9993, PhysRevD.62.024015, Kempf1995} provide a model independent way of introducing minimum length whilst recovering ordinary quantum mechanics on larger scales. GUP phenomenology has been subject to much theoretical examination \cite{Chang2002, Brau1999, Hossenfelder2006, Kempf1997, Lewis2011, Ching2013, Poutra2013, Nozari2006, Bosso2022}, yet critical insights from experiment are highly limited due to the scales at which the introduced modifications induce an observable effect.\\
The GUP may take the general form
\begin{equation}\label{eqn:GUP}
\Delta x \Delta p \geq \frac{\hbar}{2}\left[1+\beta_0\left(\frac{l_p\Delta p}{\hbar}\right)^2\right],
\end{equation}
where $\Delta x$ and $\Delta p$ denote the measurement uncertainties in the physical coordinate $x$ and corresponding conjugate momentum $p$ and $\beta_0$ is a dimensionless, model-dependent parameter that defines the minimum uncertainty implied by (\ref{eqn:GUP}); $\Delta x_\mathrm{min} = \sqrt{\beta_0}l_p$. The parameter $\beta_0$ is usually assumed to be close to unity, which leads to corrections to measurement uncertainties that only become relevant at lengths $l\sim l_p$ or energies $E\sim \frac{\hbar c}{l_p} = 1.2\times 10^9$ GeV.\\
Due to such extreme scales, direct tests of (\ref{eqn:GUP}) are challenging. However, $\beta_0$ may be constrained by measuring the perturbations to the ground state energy of a harmonic oscillator implied by (\ref{eqn:GUP}). Recently, resonant mass gravitational wave experiments, such as AURIGA exploited this effect to place upper bounds $\beta_0<10^{33}$ \cite{Marin2013}, however these bounds are still far from the expected Planck scale. Tighter bounds can be achieved by considering the application of this phenomenology to quantum mechanics.\\
Extending this formalism; one can associate a deformed Heisenberg algebra to the GUP \cite{Luis1995}
\begin{equation}\label{eqn:com}
\left[\hat{x},\hat{p}\right] = i\hbar\left(1+\beta_0\left(\frac{ l_p}{\hbar}\hat{p}\right)^2\right),
\end{equation}
the introduction of which immediately has consequences to the energy spectrum for single particle quantum systems. Further constraint on $\beta_0$ has thus been made by looking for corrections to the Lamb shift, and Landau levels of scanning tunnelling microscopes \cite{Das2008}, implied by the reconstructed quantum mechanics of (\ref{eqn:com}). However these constrains only manage an upper bound of $\beta_0<10^{20}$.\\
A different approach to excluding minimum length can be found when considering the application of the deformed Heisenberg algebra (\ref{eqn:com}) to the centre of mass modes of composite particle systems. Such systems typically exhibit larger momenta, increasing the magnitude of GUP induced effects. Interestingly, the exact implications of extending this formalism to the centre of mass mode of macroscopic systems is still unclear, with some arguments suggesting that the strength of minimum length induced effects may be suppressed by the total number of composite particles \cite{Camelia2011, Camelia2013, Kumar2020}. However, it has been remarked  \cite{Bawaj2015} that the exact constituent level at which quantum gravitational effects intervene is completely unknown. Thus, further dedicated experimental investigation into these effects, under the assumption that ($\ref{eqn:com}$) holds for the centre of mass mode of macroscopic systems, is necessary.\\
Dedicated investigations into minimum length have recently been conducted that search for macroscopic GUP induced effects such as the broadening of molecular wave-packets \cite{Villalpando2019}, various optomechanical effects \cite{Pikovski2012, Kumar2018, Bosso2017, Girdhar2020}, and perturbations to the resonant modes of mechanical resonators \cite{Marin2013,Bawaj2015, Bushev2019}.\\
\begin{figure}
\includegraphics[width=0.5\textwidth]{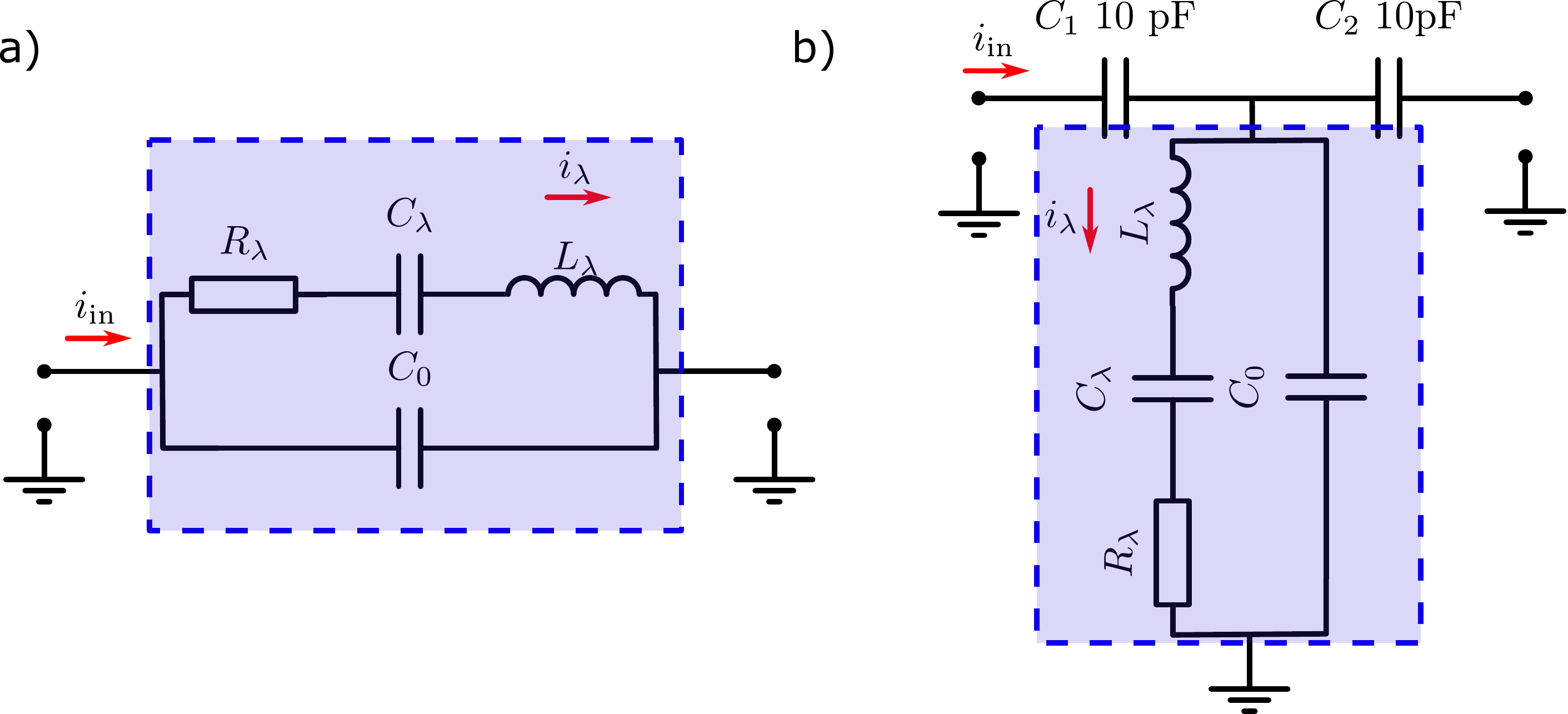}
\caption{\label{fig:schematic}Schematics of the resonator network topologies utilised. The quartz resonator in the vicinity of a resonant mode is described as a single RLC branch in parallel with the electrode capacitance as shown. The shaded blue box distinguishes this lumped element resonator model from external electrical components. a) Series network for coupling to the resonant modes $\omega_r$. b) Parallel T-network for coupling to the anti-resonant modes $\omega_a$.}
\end{figure}
In this letter, previous tests utilising mechanical resonators are improved upon. The minimum length order parameter $\beta_0$ is constrained by measurement of the non-linear frequency perturbations to the resonant modes of a macro-mechanical quartz bulk acoustic wave (BAW) resonator. Such resonators exhibit extraordinarily high acoustic quality factors at cryogenic temperatures \cite{Galliou2013, Lo2016}, and have long been used as timing standards with impressive short to mid-term frequency stability. This inherent frequency precision coupled with a gram-scale mass makes the quartz BAW resonator an ideal technology to probe minimum length induced effects, as well as performing other tests of fundamental physics \cite{Campbell2021,Goryachev2021,Goryachev2018}. The results presented in this work improve upon the upper bounds on $\beta_0$ placed by previous mechanical resonators experiments \cite{Bushev2019} by up to three orders of magnitude.\\
\section{Model}
In order to describe the perturbations to a mechanical oscillator due to the existence of a minimum length; it is commonly assumed \cite{Pikovski2012,Bawaj2015,Bushev2019} that any two conjugate position and momentum observables, attributed to the centre of mass mode of a composite system, obey the new deformed commutator of equation (\ref{eqn:com}). Whilst the form of the resonator's classical Hamiltonian remains unchanged. It then follows that a suitable representation for the physical position and momentum operators $\hat{x}, \hat{p}$ in terms of some general operators $\hat{X},\hat{P}$ can be chosen such that the canonical commutation relation $[\hat{X},\hat{P}]=i\hbar$, is recovered. One such representation can be described by the transformation \cite{Quesne2010} 
\begin{equation}\label{eqn:xp}
\hat{x},\hat{p}\rightarrow\hat{X},\hat{P}(1+\beta_0\frac{1}{3(m_pc)^2}\hat{P}^2),
\end{equation}
where $m_p$ denotes the planck mass. It has been shown \cite{Bawaj2015} that this representation leads to a non-linear correction to the resonator's Hamiltonian, resulting in a dependence of the device's perturbed resonant frequency $\tilde{\omega}_r$ on it's oscillation amplitude $x_0$:
\begin{equation}\label{eqn:omega}
\tilde{\omega}_r=\omega_r\left[1+\frac{\beta_0}{2}\left(\frac{m_\mathrm{eff}\omega_rx_0}{m_pc}\right)^2\right].
\end{equation}
Equation (\ref{eqn:omega}) immediately applies to the frequencies of the purely mechanical resonant modes of the quartz BAW resonator. However, the piezoelectric nature of the device allows for the further study of \textit{anti-resonant} modes in which an electrical half degree of freedom is contributed by the capacitive electrodes and the external circuit.\\
Application of the same GUP formalism to these anti-resonant modes can be understood by first considering the Butterworth-VanDyke model of the resonator in the vicinity of a resonant mode $\lambda$, in which the piezoelectric crystal is described by a lumped element electrical circuit consisting of a series RLC branch shunted by the capacitance of the external electrodes. An example can be seen in the shaded blue region of figure \ref{fig:schematic}. Assuming a linear coupling; the charge form equation
\begin{equation}\label{eqn:charge}
q=\kappa_\lambda x
\end{equation}
can be introduced to relate the electrical charge variable $q$ back to the mechanical displacement $x$ of the crystal plate. This allows for full interchangeability between lumped element electrical and mechanical descriptions of the resonator.\\
In order to explore the GUP phenomenology of the anti-resonant modes of this system; it is helpful to employ a Hamiltonian formalism to the electrical circuit description. This allows for GUP induced effects to be introduced in a fully consistent manner. Following the circuit quantisation procedure by Vool \textit{et al} \cite{Vool2017} the Hamiltonian of the lossless resonator can be written as
\begin{equation}\label{eqn:Hamiltonian}
\mathcal{H} = \frac{\hat{\tilde{q}}_m^2}{2 C_\lambda} + \frac{\hat{\tilde{q}}_e^2}{2 C_0} + \frac{\left(\hat{\tilde{\phi}}_m-\hat{\tilde{\phi}}_e\right)^2}{2L_\lambda}.
\end{equation}
Where $\hat{\tilde{\phi}}_i$ is the magnetic flux \textit{node} operator corresponding to the circuit node $i$ with conjugate node charge $\hat{\tilde{q}}_i$. These node operators are defined such that for a lumped element $E$ connected to nodes $i$ and $j$ the magnetic flux of that element $\phi_{E} = \tilde{\phi}_i - \tilde{\phi}_j$, with a similar definition for the conjugate charge $q_{E}$. The motional node denoted by $m$ is the intersection of the motional inductance $L_\lambda$ and capacitance $C_\lambda$, whereas the electrical node $e$ connects $L_\lambda$ and the electrode capacitance $C_0$. With the motional resistance $R_\lambda$ made to vanish; the remaining circuit node is then identified as the electrical ground $g$ such that  $\hat{\tilde{\phi}}_g,\hat{\tilde{q}}_g=0$.\\
The Hamiltonian of equation (\ref{eqn:Hamiltonian}) generates two sets of coupled equations of motion, one for each pair of variables $\hat{\tilde{\phi}}_i,\hat{\tilde{q}}_i$. Substituting these coupled equations into each other gives
\begin{equation}
\ddot{\hat{\phi}}_L+\omega_a^2\hat{\phi}_L=0,~~~~\omega_a = \sqrt{\frac{C_\lambda+C_0}{C_\lambda C_0}\frac{1}{L_\lambda}},
\end{equation}
where $\hat{\phi}_L = \hat{\tilde{\phi}}_m-\hat{\tilde{\phi}}_e$. This differential equation clearly generates oscillatory solutions with frequency $\omega_a$ corresponding to the anti-resonant modes of the circuit. Conversely, by shorting the electrical node to ground such that $\hat{\tilde{\phi}}_e,\hat{\tilde{q}}_e=0$, we recover the equations that lead to the resonant solutions;
\begin{equation}\label{eqn:seriesRes}
\ddot{\hat{\phi}}_L+\omega_r^2\hat{\phi}_L=0,~~~~\omega_r = \sqrt{\frac{1}{L_\lambda C_\lambda}}.
\end{equation}
This distinction can be understood as coupling an ideal port to the electrical node of the circuit with either infinite output impedance giving anti-resonant solutions, or zero output impedance shorting the electrical node to ground and giving resonant solutions. In a physical dissipative system, both modes can be coupled to.\\
Employing the deformed Heisenberg algebra of equation (\ref{eqn:com}) and utilising the charge form (\ref{eqn:charge}); it follows that a suitable representation for the physical charge and flux operators $\hat{q},\hat{\phi}$ in terms of some general operators $\hat{Q},\hat{\Phi}$, that recovers the canonical commutation relation $\left[\hat{Q},\hat{\Phi} \right]=i\hbar$ is
\begin{equation}
\hat{q},\hat{\phi}\rightarrow\hat{Q},\hat{\Phi}(1+\beta_0\frac{1}{3(m_pc)^2}\kappa_\lambda\hat{\Phi}^2).
\end{equation}
This representation thus perturbs the structural form of  the Hamiltonian of equation (\ref{eqn:Hamiltonian}), to leading order in $\beta_0$, by the addition of the term $\epsilon (\hat{\tilde{\Phi}}_m-\hat{\tilde{\Phi}}_e)^4/L_\lambda$, where $\epsilon = \beta_0\kappa_\lambda/3(m_pc)^2$.\\
The Heisenberg equations of motion in this new representation can be found to be 
\begin{subequations}\label{eqn:HEOM}
\begin{align}
\dot{\tilde{Q}}_m = \frac{\tilde{\Phi}_m-\tilde{\Phi_e}}{L_\lambda}\left(1+4\epsilon\left(\tilde{\Phi}_m-\tilde{\Phi_e}\right)^2\right) \\
\dot{\tilde{\Phi}}_m = \frac{\tilde{Q}_m}{C_\lambda} \\
\dot{\tilde{Q}}_e =  \frac{\tilde{\Phi}_m-\tilde{\Phi_e}}{L_\lambda}\left(1+4\epsilon\left(\tilde{\Phi}_m-\tilde{\Phi_e}\right)^2\right) \\
\dot{\tilde{\Phi}}_e = \frac{\tilde{Q}_e}{C_0} \\
\end{align}
\end{subequations}
where the hat to denote an operator has been omitted for ease of notation. Solving these coupled differential equations for $\Phi_L(t) = \tilde{\Phi}_m(t)-\tilde{\Phi}_e(t)$ via Poincare's method \cite{Strogatz2018}, and assuming $\epsilon\ll 1$, gives a solution with a perturbed anti-resonant frequency
\begin{equation}\label{eqn:omega2}
\tilde{\omega}_a=\omega_a\left[1+\frac{\beta_0}{2}\left(\frac{\kappa_\lambda \Phi_L}{m_pc}\right)^2\right].
\end{equation}
This relation can be  converted back into the mechanical description by the substitution of $\kappa_\lambda\Phi = P = m_\mathrm{eff}\omega X$. It is thus follows that under the GUP formalism the electromechanical anti-resonant modes of the quartz resonator experience the exact same degree of amplitude dependent frequency perturbation as that of the purely mechanical resonant modes described by equation (\ref{eqn:omega}). For completeness; if the electrical node is grounded such that $\tilde{\Phi}_e,\tilde{Q}_e=0$, then the solution to the resulting differential equation from (\ref{eqn:HEOM}) is that of the resonant mode with frequency given by equation (\ref{eqn:omega}).\\
Non-linear effects in quartz associated with high order elasticity terms in the crystal Hamiltonian induce a similar amplitude dependent frequency shift. This is known as the the amplitude-frequency effect and has been extensively studied in quartz resonators \cite{Tiersten1986}. As a result an observed frequency perturbation due to the deformed commutator cannot be decoupled from that due to the crystal's inherent non-linearity. However, an upper bound on $\beta_0$ can be determined up to some confidence limit by fitting equation (\ref{eqn:omega}) to any observed change in $\omega_0$ as a function of oscillation amplitude. Minimum length can thus  be constrained by measuring the frequency response of a cryogenic quartz resonator's modes under increasing oscillation amplitude.\\
\begin{figure}
\includegraphics[scale=0.5]{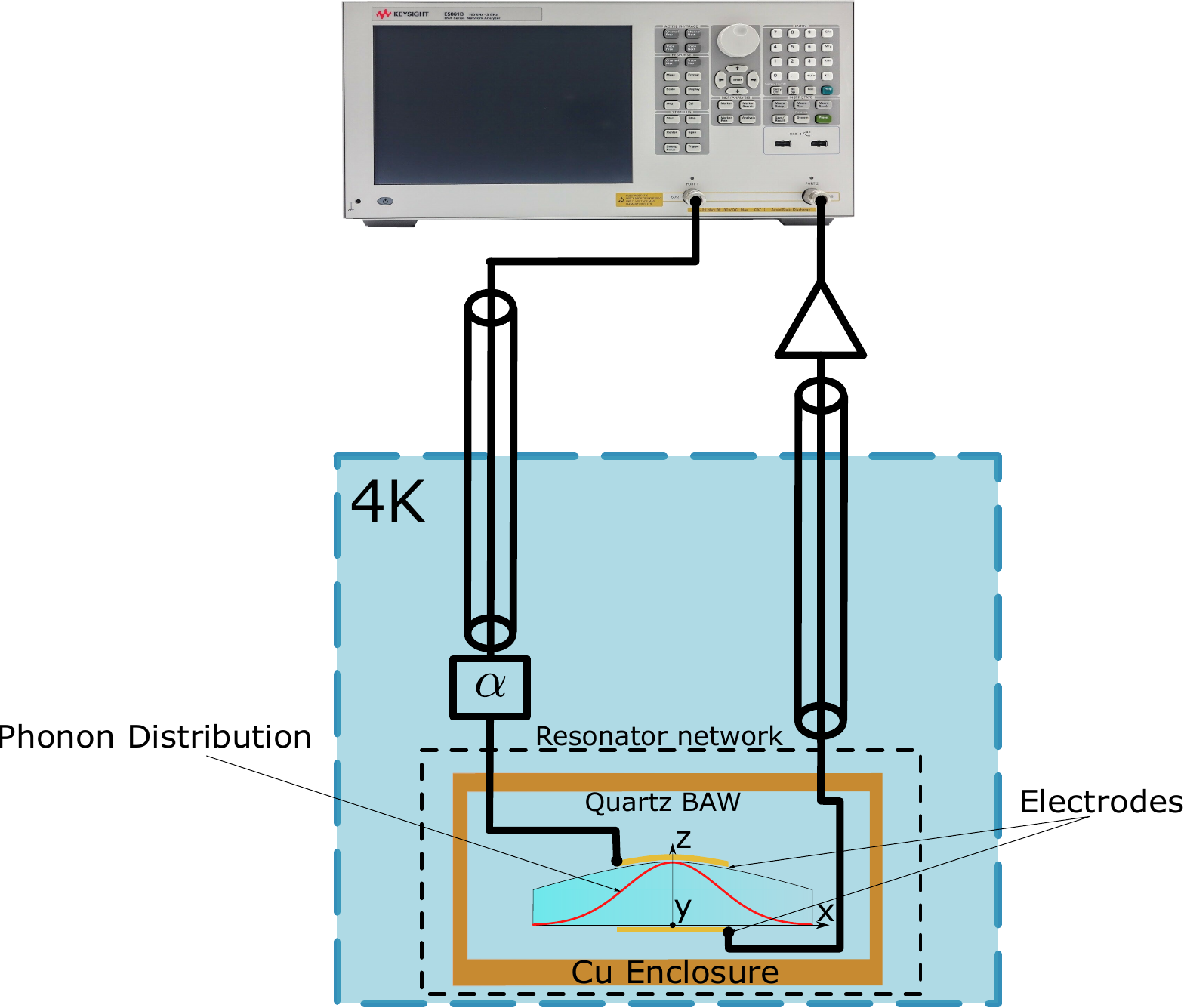}
\caption{\label{fig:setup}Diagram of the experimental set up for the case of the resonator connected in the series configuration. The loss in the input line is denoted by $\alpha$. The output signal is amplified at room temperature in order to improve the signal-to-noise of the measured acoustic modes.}
\end{figure}
\section{Methodology}
The resonator used in this work consists of a 30 mm diameter by 1 mm thick quartz crystal plate cut along the SC axis \cite{1}, with a plano-convex surface in order mitigate phonon losses due to anchoring. Copper electrodes are held close to the crystal's surface by an additional supporting quartz structure, with the entire assembly placed into a surrounding copper enclosure and sealed under vacuum pressure. A copper mount attaches the resonator assembly onto the cold plate of a dilution refrigerator, where it is subject to a stable cryogenic environment of temperature $T =4$ K. Sub-Kelvin dilution temperatures are also attainable with this set-up, however the non-linear effects in the quartz crystal become far more dominant at these extremely low temperatures \cite{Goryachev2013}. Additionally, the quality factors of certain modes may degrade for T$<$1 K. This occurs as the crystal's loss mechanisms undergo a phase transition into a regime which is dominated by two level systems associated with crystalline disorder. Thus, a higher temperature of 4 K is preferred for this work.\\
\begin{figure*}
\includegraphics[width=\textwidth]{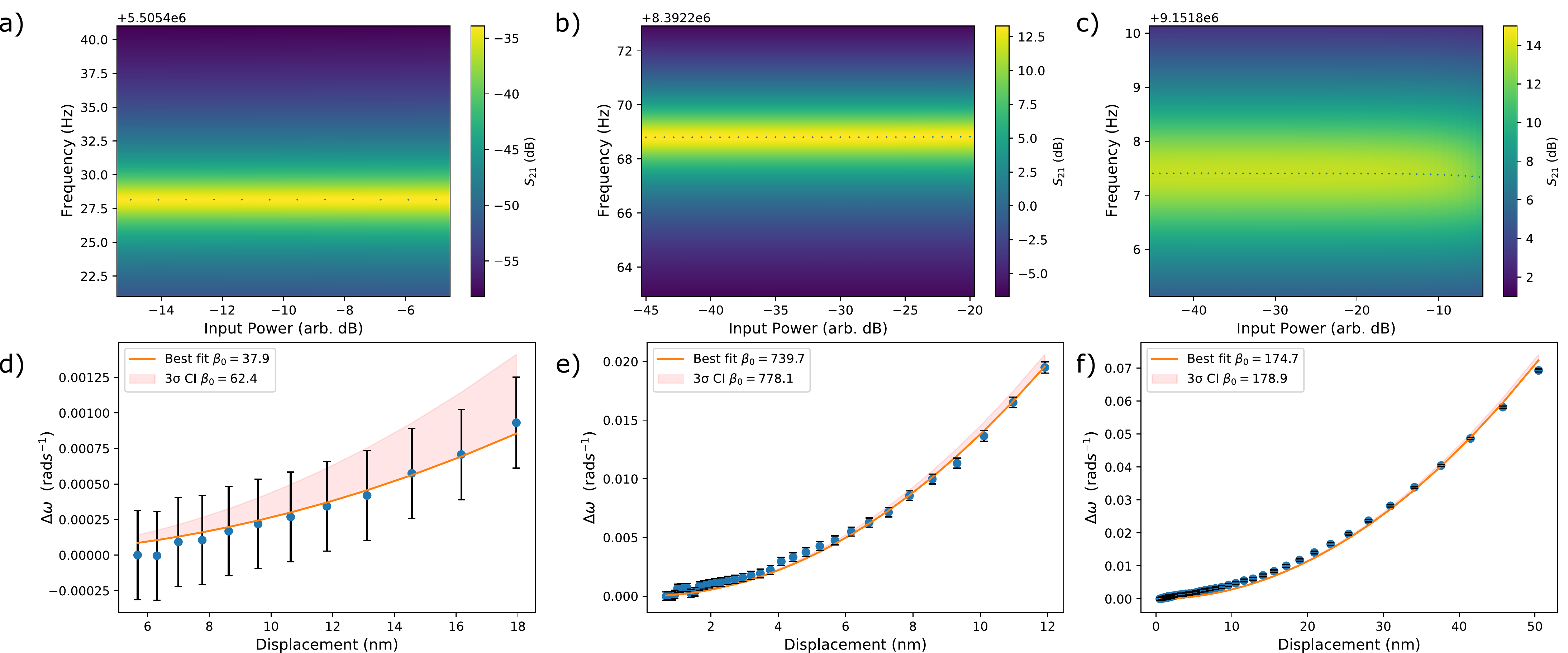}
\caption{\label{fig:fits}Plots a)-c) show the experimental results of $S_{21}$ measurements as a function of input power for the third order 5.5 MHz resonant mode, as well as the fifth order 8.4 MHz and 9.2 MHz resonant modes. The center frequency found by Lorentzian fitting has also been marked. Plots d)-f) show the net frequency perturbations $\Delta\omega$ as a function of crystal displacement amplitude, as well as the fitted limits on $\beta_0$, for the same corresponding modes.}
\end{figure*}
Two separate electrical topologies were employed in measuring frequency perturbations of the resonator in order to investigate both resonant and anti-resonant modes. Schematics of both topologies are presented in figure \ref{fig:schematic}. The motivation behind these configurations is understood by once again considering the electric circuit description of the crystal resonator. In this model; resonant or \textit{series} modes of the crystal can be described by the minimum impedance modes of the series RLC branch parametrised by the crystal's motional degrees of freedom. In addition, the anti-resonant or so called \textit{parallel} modes can be described as the minimum admittance modes of the motional RLC branch in parallel with the physical electrode capacitance. Thus, different electrical configurations can be employed to optimise coupling to either the purely mechanical resonant modes, or hybrid electromechanical anti-resonant modes of the resonator.\\
The first topology utilised was a series configuration where each of the resonator's electrodes were coupled to individual co-axial signal lines that feed out of the cryogenic system. This allows for an electrical drive signal supplied by external source to excite resonant acoustic modes in the crystal bulk, due to the piezoelectric coupling of the quartz lattice to a charge density on the electrodes. The same mechanism can be exploited in order to read out the mechanical motion of the crystal plate.\\
In the parallel configuration, the quartz electrodes were directly soldered to a capacitive T-network. This circuit was then coupled to the signal lines in such a way as to shunt the quartz resonator to the electrical ground plane. This configuration allowed for direct coupling to the anti-resonant modes of the resonator, through the same piezoelectric mechanism. Studying the anti-resonant modes of this system is of interest as the resonance shares a half degree of freedom with the purely electrical component in the parallel capacitance, perhaps modifying the degree of frequency perturbation due to inherent crystal non-linearities.\\
Impedance measurements of the resonant network were made utilising a vector network analyser (VNA) locked to a stable frequency reference supplied by Hydrogen Maser, with signals supplied by co-axial cables that feed into the cryogenic system and connect to the resonators electrodes. The measurement set up is displayed diagrammaticality in figure \ref{fig:setup}. Measurements of the system transmission parameter $S_{21}$ as a function of frequency $\omega$ in the vicinity of a mode, show maximum transmission at the frequency corresponding to the resonant mode $\omega_r$ for the case of the series configuration, or the anti-resonant mode $\omega_a$ for the case of the parallel T-network configuration. The exact frequency of each mode, as well as their quality factors $Q_\lambda$, can thus be determined by fitting a Lorentzian to the corresponding $S_{21}(\omega)$ measurements.\\
\begin{table*}
\begin{tabular}{c||c|c|c|c|c|c|c|c}
\hline\hline
$\lambda = X_{n,m,p}$ & $\omega_{r}/2\pi$ (Hz) & $\omega_{a}/2\pi$ (Hz) & $Q_\lambda$ ($10^6$) & $R_\lambda$ ($\Omega$) & $m_\mathrm{eff}$ (mg) & $\kappa_\lambda$ ($10^{-4}$C/m) & $\beta_{0,3\sigma}^{(r)}$ & $\beta_{0,3\sigma}^{(a)}$\\
\hline
$\mathrm{B}_{3,0,0}$ & 5 505 406 & 5 505 613 & 54.7 & 2.9 & 8.5 & 13.6 & 62.4 & 27 971 \\
$\mathrm{C}_{5,0,0}$ & 8 392 263 & 8 392 296 & 106.5 & 5.36 & 7.37 & 8.27 & 777.8 & 13 190 \\
$\mathrm{B}_{5,0,0}$ & 9 151 797 & 9 151 853 & 60.5 & 5.9 & 6.04 & 9.87 & 178.9 & 4 426\\
\hline
\end{tabular}
\caption{\label{tab:results}Results from the investigations of resonant and anti-resonant quartz BAW modes. Each mode is denoted by $\lambda = X_{n,m,p}$ where $X$ is the polarisation ($A$ for longitudinal, $B$ and $C$ for fast and slow shear polarisations), and integers $n,m,p$ denote the longitudinal and x-y plane wave numbers respectively. $\beta_{0,5\sigma}^{(r)}$ and $\beta_{0,5\sigma}^{(a)}$ are the excluded values of $\beta_0$ to 5$\sigma$ confidence derived from fitting to the resonant and anti-resonant modes respectively.}
\end{table*}
The VNA supplies a power $P_\mathrm{VNA}$ to the system through the signal input line that drives the acoustic modes. The transfer function that relates $P_\mathrm{VNA}$ to the crystal displacement $x$ in the vicinity of resonance can be written as:
\begin{equation}\label{eqn:transfer}
x(\omega) = \frac{1}{\omega\kappa_\lambda}\left|\sqrt{\frac{(1-\Gamma)(P_\mathrm{VNA}+\alpha)}{\mathrm{Re}~[Z_\mathrm{input}(i\omega)]}}Z_\mathrm{gen}(i\omega)\right|
\end{equation}
Where $\Gamma$ is the reflection coefficient of the resonant mode, $\alpha$ is the loss in the input line, $Z_\mathrm{input}(i\omega)$ is the complex input impedance of the resonator network and $Z_\mathrm{gen}(i\omega)$ is a generalised transfer function that relates the current input to the resonator network to the  motional branch current $I_\lambda$. The coupling constant $\kappa_\lambda$ is experimentally determined and has units of C/m.\\
The term under the square root sign in equation (\ref{eqn:transfer}) represents the total current at the input of the resonator network as $I_\mathrm{input}^2=P_\mathrm{input}/Z_\mathrm{input}$ where $P_\mathrm{input}=(1-\Gamma)(P_\mathrm{VNA}+\alpha)$. The term $Z_\mathrm{gen}(i\omega)$ is then the ratio of impedances that form the current divider equation which determines $I_\lambda$ from the input current. This ratio is different for each resonator network topology. The values of $\Gamma$ and $\alpha$ for each resonant mode are experimentally determined from $S_{11}(\omega)$ measurements of the resonator network in reflection.\\
\section{Results}
\begin{figure}
\includegraphics[width=0.5\textwidth]{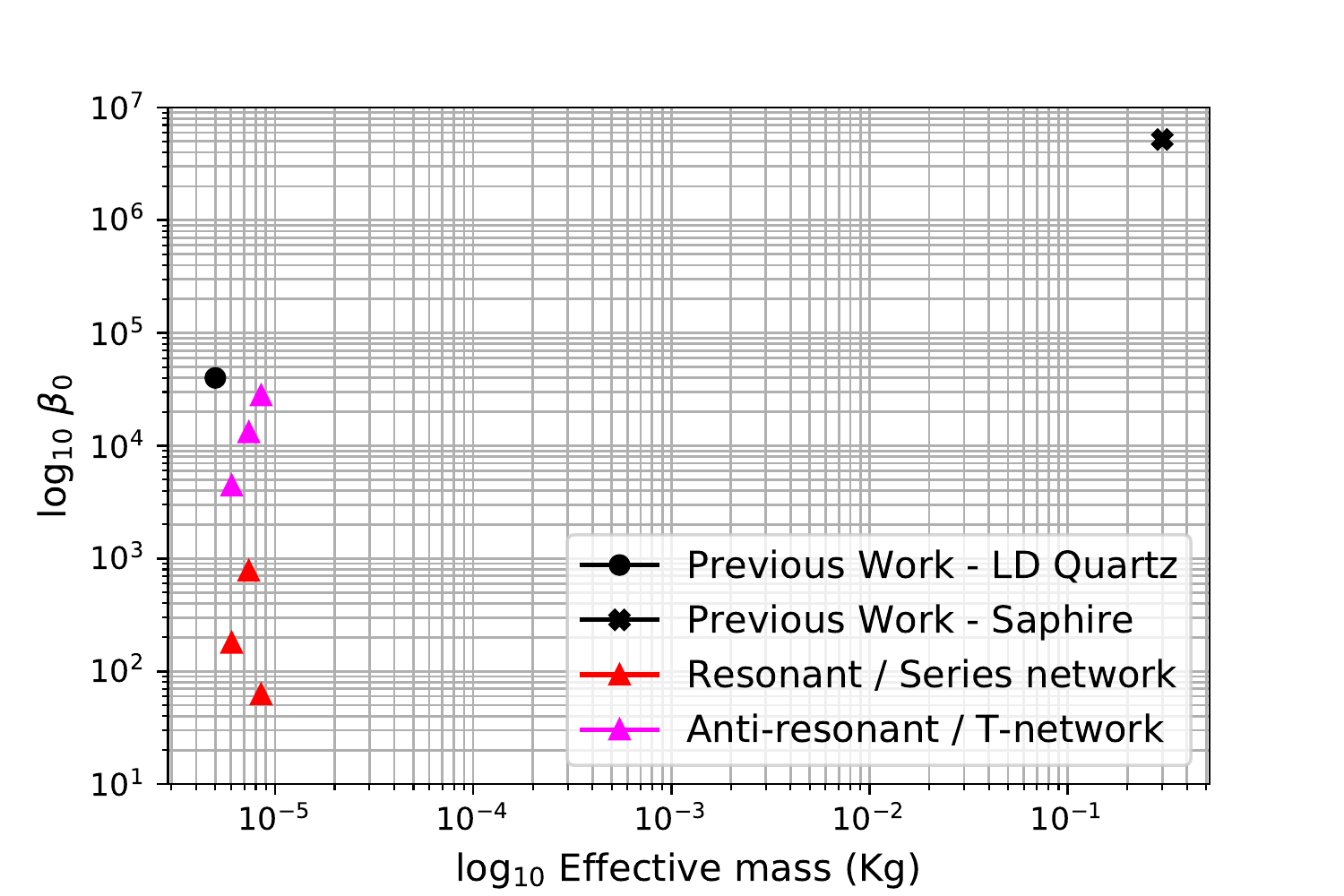}
\caption{\label{fig:exec plot}Exclusion limits from this work are plotted as well as the previous best limits generated by the sapphire mechanical resonator of ref. \cite{Bushev2019}. Also shown is a previously projected estimate for the sensitivity attainable with an LD-cut quartz resonator.}
\end{figure}

To utilise equation (\ref{eqn:transfer}) the values of the motional resistance $R_\lambda$, capacitance $C_\lambda$, inductance $L_\lambda$ and electrode capacitance $C_0$ for each mode must be known in order to construct $Z_\mathrm{input}$ as well as $Z_\mathrm{gen}$. Accurate measurements of $R_\lambda$ for multiple modes have already been conducted in previous works \cite{Goryachev2014}. All other parameters can be determined by solving the analytical expressions that minimise $Z_\mathrm{input}$ for resonant modes or $1/Z_\mathrm{input}$ for anti-resonant modes, given the experimentally determined values for $\omega_r$, $\omega_a$ and $Q_\lambda$.\\
In order to vary the excitation amplitude; $P_\mathrm{VNA}$ was increased in increments to the maximum value attainable before the excessive power input causes the cryogenic environment to increase in temperature. At each stage the $S$-parameter measurements were made with a narrow resolution bandwidth in order to avoid any ring-down effects. With the modal frequencies determined at each step, equation (\ref{eqn:omega}) can be fit to the resulting data with $\beta_0$ set as a free parameter. The errors in the center frequencies $\omega_{r,a}$ associated with the Lorentzian fitting are accounted for by running monte-carlos simulations in which $\beta_0$ is fit for multiple Gaussian distributed trajectories. Confidence limits to $3\sigma$ on the fitted values of $\beta_0$ are then obtained from the aggregate statistics. Examples of such fitting of multiple resonant modes are presented in figure \ref{fig:fits}, with the full list of all results and measured parameters given in table \ref{tab:results}, and final exclusion limits plotted in figure \ref{fig:exec plot}.\\
In this work only fundamental longitudinal overtone modes where investigated as they typically exhibit higher quality factors, and larger effective mode mass. In general the resonant or series modes provided better exclusions on the minimum length order parameter $\beta_0$ than the anti-resonant modes. This is potentially due to the external circuit capacitors of the T-network contributing extra losses and non-linear behaviours when coupled to the resonator.  The strongest limit established to 3$\sigma$ confidence was $\beta_0<62.4$, this was achieved with the 3rd order fast shear mode $\mathrm{B}_{3,0,0}$ at 5.505 MHz. When compared to previous experiments that exclude minimum length with mechanical resonators, the previous most stringent limits where set by a similar quartz resonator system at room temperature \cite{Bushev2019}. We improve upon these bounds by three orders of magnitude thanks to the improved accuracy of the analysis in which equation (\ref{eqn:omega}) is fit to experimental data, as well as the operation of the resonator in a cryogenic environment.\\ 
\section{Discussion}
It is acknowledged that tight bounds of $\beta_0 \ll 1$ have been proposed \cite{Kumar2020} by reusing results from dated experiments measuring the period of heavy pendulums \cite{Smith1964, Fulcher1967}. However, the lack of metrological treatment in these experiments limits the quantitative significance of these results \cite{Bushev2019}. For genuine validation of these bounds such experiments would need to be repeated in a controlled context; where system frequency stability, external environmental influences and spurious mode induced non-linearities can be measured.\\
Multiple discussions around the physical conclusions that can be drawn from experimental tests of minimum length remain highly active and seemingly problematic. As mentioned above; the application of the GUP to center of mass coordinates is uncertain. In the case that GUP induced effects in macroscopic bodies are to be suppressed by either particle number or the number of elementary interactions, the results presented in this work would be scalable and remain of large significance as the most stringent metrological bounds to date. It has also been noted \cite{Bawaj2015} that the phenomenological model which leads to (\ref{eqn:omega}) suggests large deviations from the classical trajectories of astronomical objects, possibly leading to conclusions such as $\beta_0$ displaying some mass dependence. In this case it becomes well motivated to explore minimum length over a varying range of mass. The quartz BAW resonator thus presents an attractive platform for such explorations thanks to a large density of overtone modes giving access to the effective mass regime of a few milligrams and below. \\
An additional complication arises as exclusion of $\beta_0$, based on the model of equation (\ref{eqn:xp}), approaches bounds of $\mathcal{O}(1)$. The chosen representation of the physical position and momentum operators is a perturbative approach that recovers minimum length for $\beta_0\gg1$. A more exact nonperturbative treatment is therefore necessary  for smaller length scales \cite{Bosso2022}. As the results presented in this work investigate $\beta_0$ at an order of magnitude larger than $\beta_0 = 1$; this perturbative approach is sufficient. Further to this point, the conclusions that should be drawn by these results are an exclusion bound on the potential scale of minimum length intervention in macroscopic systems, and not any sort of confirmation of the GUP approach or phenomenology.\\
These arguments motivate the need for further experimental investigation into the behaviours of center of mass modes of macroscopic bodies close to the quantum regime. Exploring the classical to quantum transition in mechanical resonators with minimal decoherence to the external environment could reveal further insights into GUP implications of composite systems as well as other quantum phenomena such as spontaneous wave function collapse \cite{Lajos2015, Vinante2020, TobarG2022}. Previous work has been undertaken in bringing low noise nano-mechanical resonators into the quantum regime through various cooling procedures \cite{OConnell2010, Teufel2011, Chan2011, Cattiaux2021, Naseem2021, Seis2022}, setting the stage for the confirmation of purely quantum phenomena in composite particle systems \cite{Amir2012}. Further development of quantum limited mechanical devices of even larger mass is ongoing \cite{Georgescu2022}. Thus, pushing the quartz BAW platform to single phonon occupancy, which would be attainable for $\omega > 200$ MHz in a 10 mK thermal environment, is an ideal next step towards the demarcation of the quantum-classical limit.\\
\section{Conclusions}
In summary; minimum length has been tested by exploring the effect of the GUP on the center of mass mode of a macroscopic quartz BAW resonator. Constraints on the minimum length order parameter $\beta_0$ have been placed to $3\sigma$ confidence by fitting to the observed frequency perturbations of the resonant and anti-resonant modes of several crystal resonances. The constraints derived in this work improve upon the previous most stringent limits obtained with dedicated metrological systems by at least three orders of magnitude, thanks to an improved analysis technique as well as a low noise cryogenic operating environment. As further bounds on $\beta_0$ by macroscopic resonators approach the theoretically predicted regime of $\beta_0=1$, stringent approaches and analysis will become necessary in order to draw physical conclusions. Further exploration of the quantum regime in macroscopic acoustic systems, such as single-phonon mechanical resonators, will hopefully provide the necessary developments in order to make conclusive statements on the GUP phenomenology, as well as other testable predictions of quantum theories.

\bibliography{QGBAW}
\end{document}